\documentclass[twocolumn,APS]{revtex4}

\def \and{\& }

\def\RefAIP#1{\noindent\hangafter=1\hangindent=0.25in[#1]}
\def\by#1{#1,}
\def\and{and }
\def\yr#1{(#1)}
\def\paper#1{}
\def\jour#1{#1}

\def\vol#1{{#1}}
\def\issue#1{}
\def\pages#1{\hbox{#1.}}

\def\JFM      {J.~Fluid Mech.\ }

\usepackage{amsmath}
\usepackage{amsfonts}
\usepackage{amssymb}

\usepackage{graphicx}

\begin{document}

\title{Superfast non-linear diffusion: Capillary transport in particulate porous media.}
\bigskip

\author{A.V. Lukyanov$^1$, M.M.  Sushchikh$^2$, M.J. Baines$^1$, T.G. Theofanous$^2$}
\medskip

\affiliation
{Department~of~Mathematics and Statistics,~University~of
Reading, ~Reading~RG6~6AX,~U.K.$^1$}

\affiliation
{Center for Risk Studies and Safety,~University~of
California at Santa Barbara, Santa Barbara, CA, USA.$^2$}
\bigskip

\begin{abstract}
The migration of liquids in porous media, such as sand, has been commonly considered at high saturation levels with liquid pathways at pore dimensions. In this letter we reveal a low saturation regime observed in our experiments with droplets of extremely low volatility liquids deposited on sand. In this regime the liquid is mostly found within the grain surface roughness and in the capillary bridges formed at the contacts between the grains. The bridges act as variable-volume reservoirs and the flow is driven by the capillary pressure arising at the wetting front according to the roughness length scales. We propose that this migration (spreading) is the result of interplay between the bridge volume adjustment to this pressure distribution and viscous losses of a creeping flow within the roughness. The net macroscopic result is a special case of non-linear diffusion described by a {\it superfast diffusion equation} (SFDE) for saturation with distinctive mathematical character. We obtain solutions to a moving boundary problem defined by SFDE that robustly convey a time power law of spreading as seen in our experiments. 
\end{abstract}

\maketitle

Extremely low volatility (persistent) liquids can spread significantly in porous substrates and wet large volumes over long periods of time before they are removed by evaporation. For example, we found in experiments that a $2.8 \,\mu\mbox{L}$ drop of Tris(2-ethylhexyl) phosphate (TEHP - vapour pressure $\sim10^{-5} \,\mbox{Pa}$) deposited on a bed of Ottawa sand, covered a volume of $\sim1.7\,\mbox{mL}$ in $25$  days, where the spreading essentially came to a stop. With a measured (wet-bed) porosity  of  $\phi\sim30\%$, this suggests an average saturation level of $\bar{s}\sim0.55\%$ ($s$ is the liquid volume divided by the volume of available pore space) and, if uniformly distributed on the surface of $250\,\mu\mbox{m}$ spherical grains, a "film" thickness of $\sim100\,\mbox{nm}$. Clearly we are confronted with the mechanics of a process unlike anything found in previous investigations of liquid spreading in environmental porous media. 

\begin{figure}
\begin{center}
\includegraphics[trim=0cm 0cm 0cm 0cm,width=0.45\columnwidth]{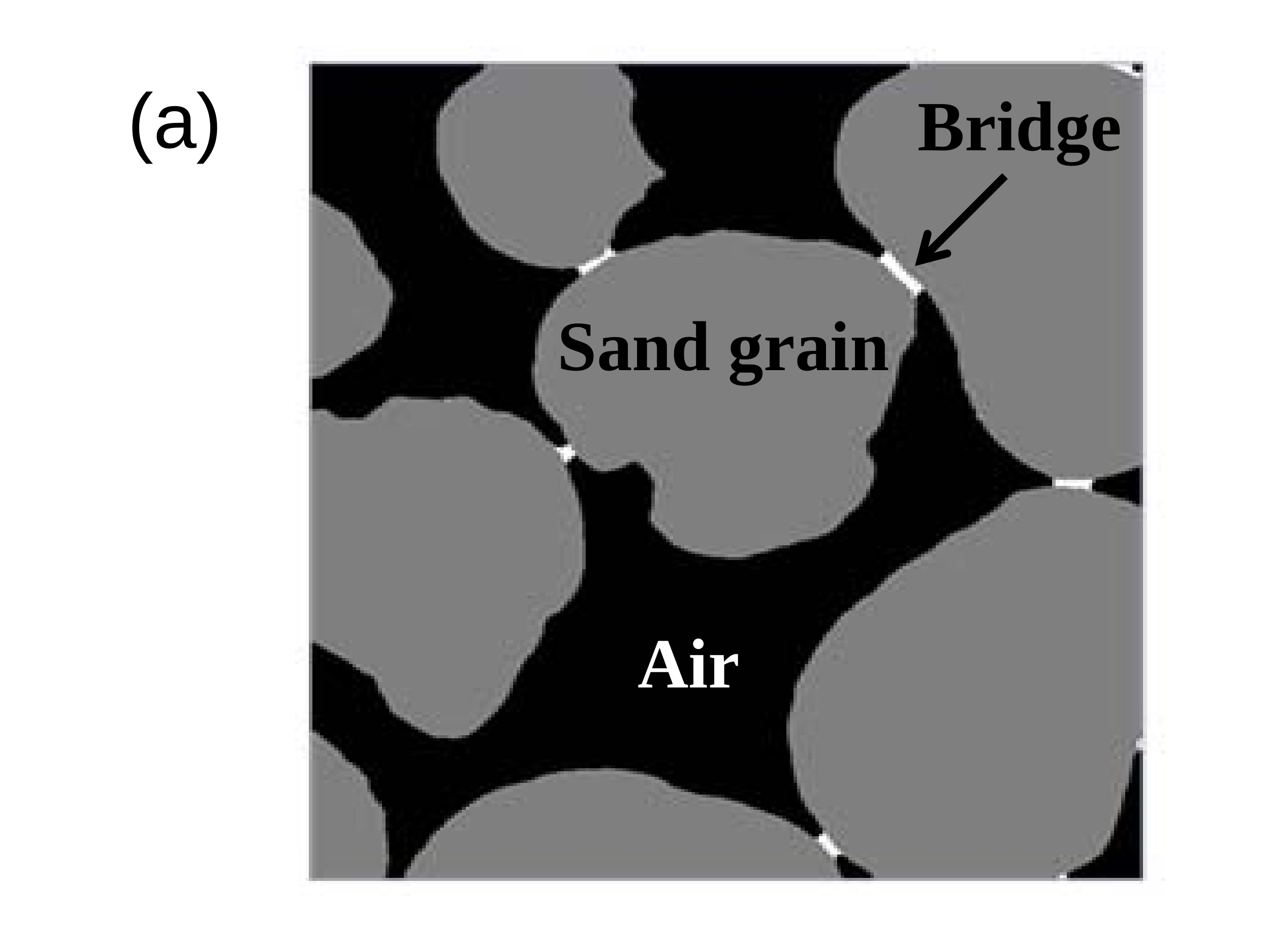}
\includegraphics[trim=0cm 0cm 0cm 0cm,width=0.45\columnwidth]{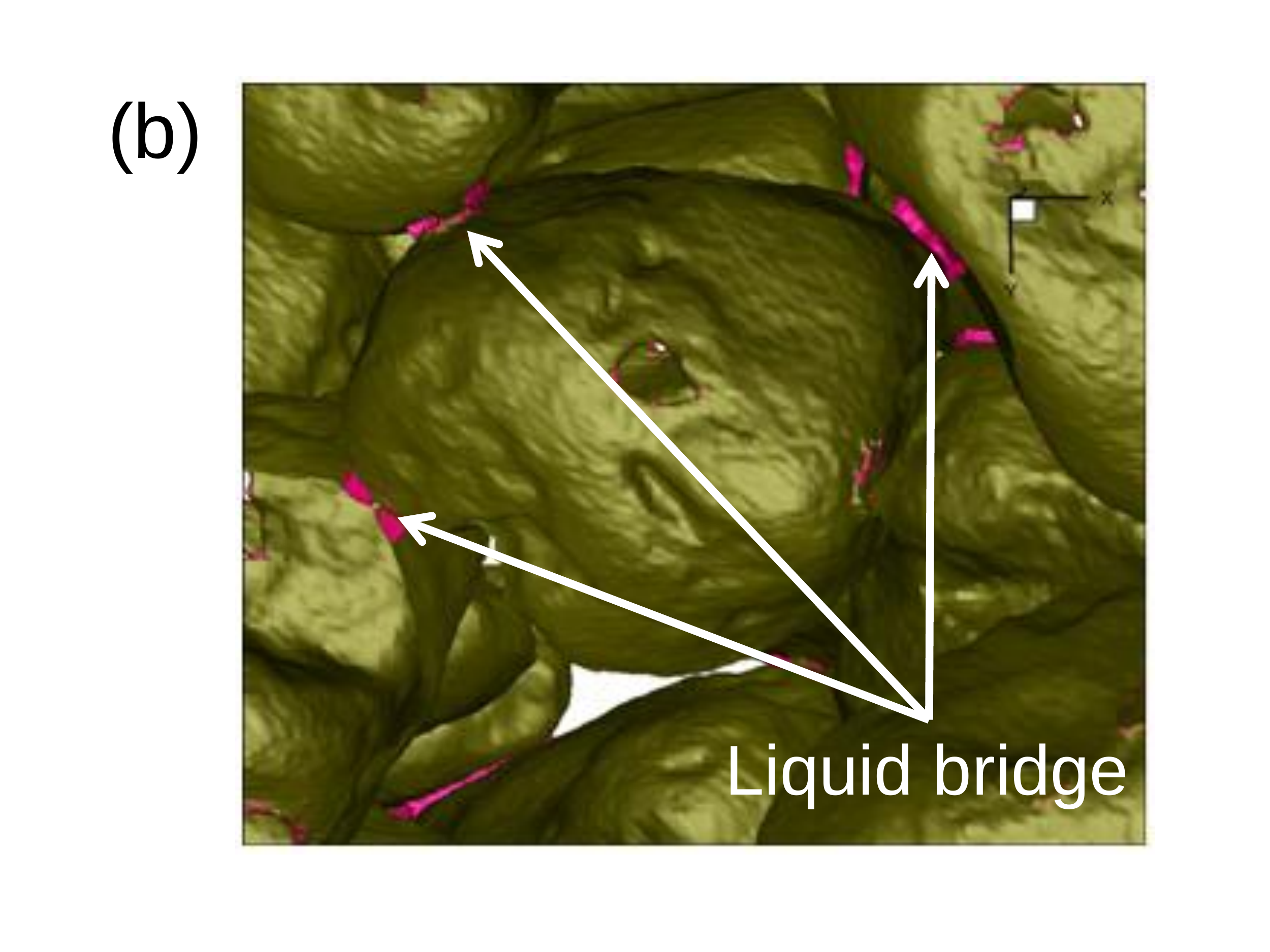}
\end{center}
\caption{Illustration of isolated bridges at low levels of saturations. (a) Micro X-ray Computer Tomography (MicroXCT) image, typical from our experiments. (b) 3D image reconstruction of MicroXCT data. The liquid within the grain roughness is invisible to MicroXCT, since resolution is limited to $2\,\mu\mbox{m}$.} \label{Fig1}
\end{figure}

Equilibrium considerations at low saturation levels have been made in connection with structural properties of granular materials (sands, soils), and these provide a useful starting point [1--3]. The experiments conducted with spherical glass particles ($240-1200\,\mu\mbox{m}$, unknown roughness), shaken with water and allowed to settle, have revealed that liquid domains may exhibit different morphologies. At the low end of saturations $\bar{s}_{min}<\bar{s}<8\%$ the liquid forms isolated pendular rings (liquid bridges) at the points of particle contact. The minimal saturation at the onset of bridge formation was found to be $\bar{s}_{min}\sim0.2\%$ [3], which is also the level at which the system loses its cohesive property. At higher saturations, $8\%<\bar{s}<24\%$, some bridges coallesce into more complex structures (trimers, pentamers and heptamers), but still global connectivity at the pore scale is lacking. [We have observed similar behavior with sand particles at $\bar{s}<20\%$ (Fig. \ref{Fig1}).] 

We anticipate that these overall guidelines for loss of pore-scale connectivity also apply to non-equilibrium systems, such as in our experiment described above, thereby deducing that any transport observed at such low levels of saturation must be due to capillary action at the surface-roughness scale of individual grains (a web-like surface flow geometry). This idea is pursued here by means of a relatively simple mathematical model that ties together the key physics of this transport phenomenon  and by highly sensitive experiments devised specifically for accurate tracking of wetting fronts at such trace-quantity conditions.

Previous work with liquids spreading along microscopic surface grooves of various shapes and dimensions has shown that the flow obeys a local Darcy-like law with the permeability related to the groove geometry [4]. Taking this as an approximate model of the micro-flow within the surface roughness, applying intrinsic liquid averaging $<...>^l=V_l^{-1}\int_{V_l}...\, dV$ and the spatial averaging theorem [5], one can deduce that the macroscopic, averaged over the effective continuum liquid flux $\bf q$ and pressure $p$ also obey a Darcy-like relationship   
\begin{equation}
{\bf q}=\frac{S_e}{S} <{\bf u}>^l=-\frac{\kappa}{\mu} \frac{S_e}{S} \nabla <\psi>^l=-\frac{\kappa}{\mu} \frac{S_e}{S} \nabla p.
\label{New1}
\end{equation}
Here, $\bf u$ and $\psi$ are the liquid velocity and pressure averaged within a surface ``groove'', $\mu$ is viscosity, $S_e$ is the area of entrances and exits for the liquid phase through the macroscopic surface element $S$, and $\kappa$ is the individual-groove permeability. The latter, if modified by the surface ratio $S_e/S$, yields an effective porous medium permeability $K$. An order of magnitude estimate of this quantity can be obtained from the geometry, roughness amplitude $\delta$ and the assumption that the wetted grooves are completely filled [4,6], namely, $K=\kappa \frac{S_e}{S}\approx \frac{\delta^2}{32\pi} \frac{S_e}{S} \approx  3\,(1-\phi)\frac{\delta ^3}{32\pi R}$, where $R$ is the effective particle radius. Similarly, the saturation due to the liquid within the roughness is $s_r\approx\frac{1-\phi}{\phi}\,\frac{3\delta}{2R}$.

The connection of (\ref{New1}) to the total saturation $s$ is made by accounting for the liquid content in the bridges $s_b=s-s_r$. Based on the results of [7], we obtain, for the case of two identical spheres in contact and assuming near-complete wetting systems, an approximate expression with maximum error $<1\%$ for the normalized mean curvature $HR$ as a function of the normalized bridge volume $V_b R^{-3}$, 
$HR=C_0 - C_1 (V_b R^{-3})^{\gamma}$ with $C_0=3.7$, $C_1=1.3$ and $\gamma=-0.516\pm0.001$. The saturation due to the liquid in the bridges is $s_b=\alpha^{-1}V_b R^{-3}$, $\alpha=\frac{4\pi}{3N}\frac{\phi}{1-\phi}$ and $N$ is the particle coordination number. Since $p=2\sigma H$, one obtains 
\begin{equation}
\label{CapillaryPressure}
p(s)=\frac{2\sigma}{R} \left\{ C_0 -C_1 \, \alpha^{\gamma} (s-s_r)^{\gamma}\right\},
\end{equation}
where $\sigma$ is surface tension.

\begin{figure}
\begin{center}
\includegraphics[trim=0cm 0cm 0cm 0cm,width=0.7\columnwidth]{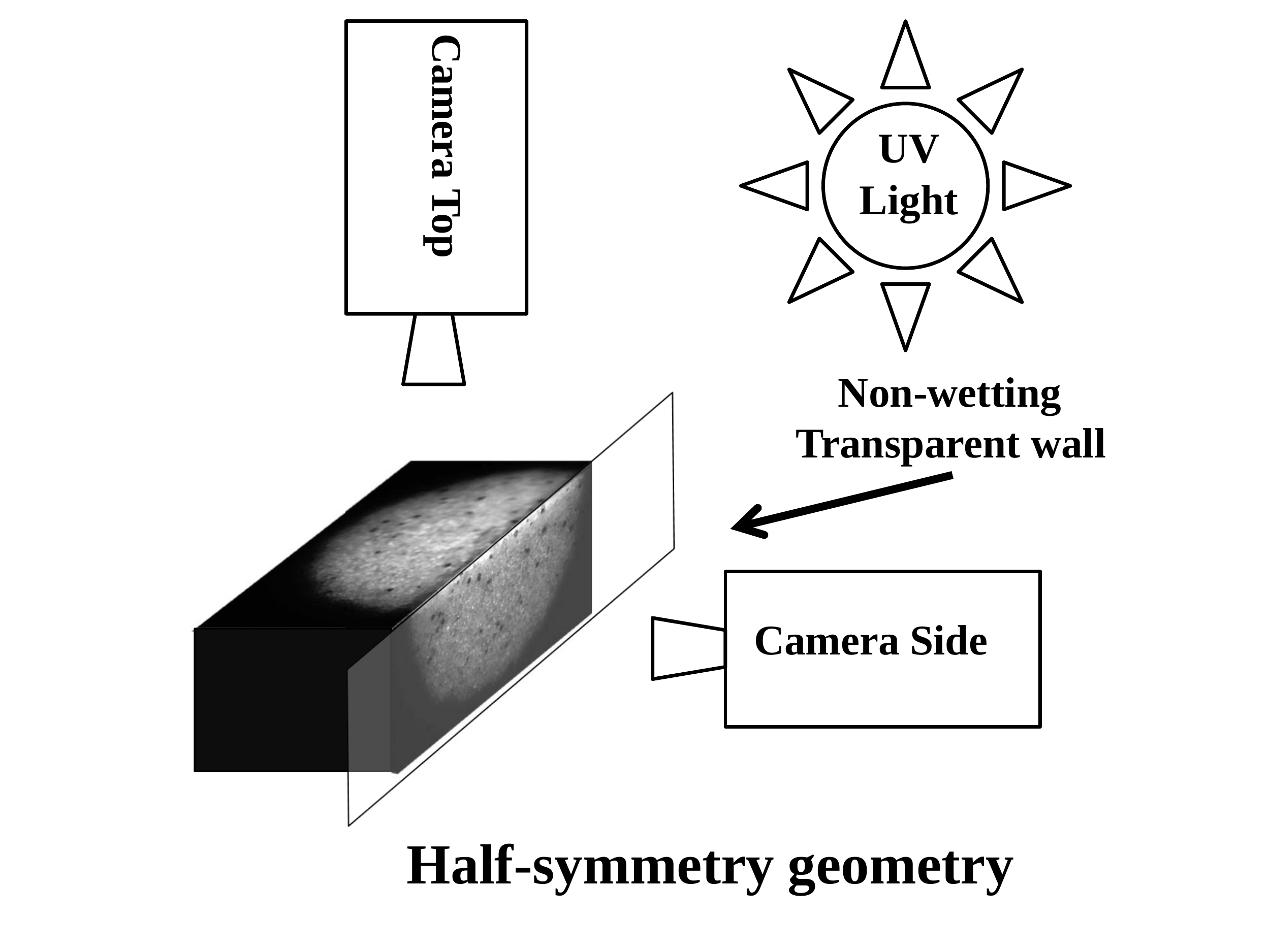}
\end{center}
\caption{ Illustration of the experimental setup for the half-symmetry tests.} \label{Fig2}
\end{figure}

Now using (\ref{New1}) and (\ref{CapillaryPressure}), one can cast the continuity equation, $\frac{\partial \phi s}{\partial t}+\nabla\cdot {\bf q}=0$, into a non-linear diffusion equation which, after introducing non-dimensional variables  $\tilde{s}=(s-s_r)/s_{m}$, $\tilde{\bf x}={\bf x}/L$ and $\tilde{t}=t/\tau$ with $\tau=L^2 s_{m}^{1-\gamma}/D$, takes the form
\begin{equation}\label{NLDiff}
\frac{\partial \tilde{s}}{\partial \tilde{t}}=\nabla\left( {\tilde{s}^{\gamma-1}}\nabla\tilde{s}\right),
\end{equation}
where $s_{m}$ and $L$ are the characteristic saturation and the length scale of the system and $D=\frac{2\sigma}{R}\frac{K }{\mu }\frac{C_1 |\gamma| \alpha^{\gamma}}{\phi}$ is the diffusion coefficient. 

It can be readily seen that the character of this diffusion problem depends utterly on the type of non-linearity dictated by the value of  $\gamma$. If $\gamma>1$, we have what is known as the Porous Medium Equation (PME) [8] - the typical modelling paradigm in the area of porous transport. Diffusivity in the case of PME yields zero flux at the edge of any initial condition with compact support. This in turn leads to the so-called waiting times and infinite slopes in solutions at moving boundaries, which require special numerical treatment based on a moving-mesh method [9].  If, on the other hand, $\gamma<0$, which is the case here, the diffusivity, and therefore the flux, become arbitrarily large as $\tilde{s}_F=\frac{s_F-s_r}{s_m}\to0$; this is a manifestation of the capillary pressure singularity in (\ref{CapillaryPressure}). In this case, one must have non-zero saturation at the front $\tilde{s}=\tilde{s}_F>0$ for a solution to a boundary problem to exist. Accordingly, we must supplement (\ref{NLDiff}) with the front velocity, which from the flux continuity at the moving front with a normal vector $\bf n$, in the dimensionless form reads 
\begin{equation}
\tilde{v}_F=-\frac{s_m}{s_F} \tilde{s}_F^{\gamma-1}\,({\bf n}\cdot\nabla  \tilde{s}).
\label{frontV}
\end{equation}

A mathematical discussion of this special class of problems can be found in [8] under the term {\it superfast diffusion} (SFD). 
In our case parameter $\gamma=-0.5$ is fixed and any solution to (\ref{NLDiff})-(\ref{frontV}) belongs to a two-parameter family of functions depending on $s_F/s_m$ and $\tilde{s}_F$. Since $s_F$ is the terminal saturation, its value expresses the final extent of the wet volume. The second parameter $\tilde{s}_F$ is the measure of capillary suction at the front - the larger $\tilde{s}_F$, the smaller the capillary pressure. The structure of this solution space is illustrated after the presentation of the experimental data, further below. 

Our experiments covered a variety of sands and several well-wetting, low-volatility (organophosphate) liquids  of varying viscosities and surface tensions ( Tributyl phosphate (TBP), CAS 126-73-8; TEHP, CAS 78-42-2; and Tricresyl Phosphate (TCP), CAS 1330-78-5; Sigma-Aldrich). The liquids have viscosities  $3.5/\,15/\,20\,\mbox{mPa}\cdot\mbox{s}$ at $20^{\circ}\,C$  [10], surface tensions, measured in our laboratory at $25^{\circ}C$, $28/\,29/\,42.5\,\mbox{mN}/\mbox{m}$ and contact angles measured on smooth/rough glass $10^{\circ}/0^{\circ}; 10^{\circ}/0^{\circ}; 30^{\circ}/20^{\circ}$ respectively. Initial tests showed that spreading is slowed down in the presence of re-entrant microscopic cavities on the sand grains as found for example in erosion sands - they act as irreversible liquid reservoirs. Therefore, it is sufficient for our purposes here to focus on the well-characterized (for example, [11]) Standard Ottawa sand (EMD Chemicals, p/n SX0075). The sand has an average grain size of $250\,\mu\mbox{m}$ and surface roughness in the range $0.25<\delta<3\,\mu\mbox{m}$ [11]. Natural packing (slight shaking to level the bed) yields porosities of $\sim35\%$, but when wetted we measured (by analysis of images such as in Fig.1)  $\phi\sim30\%$, and a coordination number  $N\sim7-8$. The experiments were carried out by depositing gently (fall distance $\sim5\,\mbox{mm}$) a liquid drop and following the evolution of the wet region by imaging the top surface of the bed. In addition, penetration into the bed was observed in half-symmery tests as illustrated in Fig. \ref{Fig2}. The time-lapse photography was computer-controlled in a setting that was left undisturbed inside a dark-curtain envelope over periods of many days at controlled temperature. Variability was checked by testing two side-by-side drops placed far enough to avoid interference during spreading, as well as by repeated tests with single drops of two different sizes.

The crucial part of testing was ensuring that the position of the spreading front was captured reliably both in the images and in their analysis. Proper visualization was achieved by UV-excited fluorescence; $1\%$ Coumarin $503$ dye, liquid properties confirmed to be unaffected. Photographs were taken by two $10.7$ MP b/w digital cameras (Lumenera Corporation) equipped with macrolenses and focused to resolve individual grains.The lenses were fitted with longpass glass filters to cut off scattered excitation light. No significant background signal could be detected in the absence of the dye in the range of exposures used in the experiments. We have taken images at different values of the exposure to ensure confident capture of the wetted front position  at saturation levels down to $0.1\%$ (mixed-to-equilibrium sand). In analysis, a threshold level of signal at the wetted front was clear and used in automatic scanning of images to determine wetted areas as a function of time. 

\begin{figure}[htp]
\begin{center}
\includegraphics[trim=0cm 0cm 0cm 0cm,width=0.9\columnwidth]{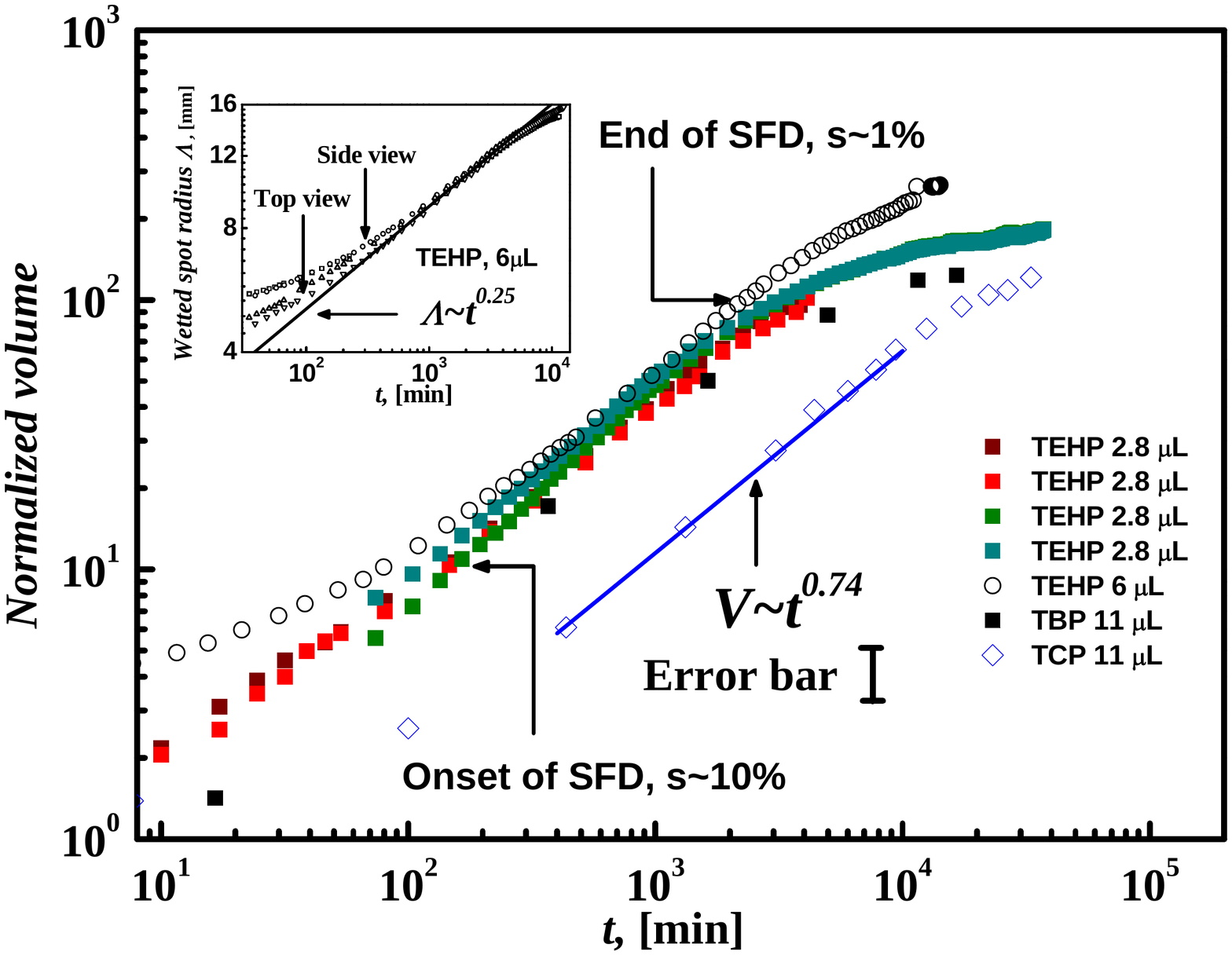}
\end{center}
\caption{ Spreading of various liquids and drop quantities in sand at seven full-symmetry tests; inset - two half-symmetry tests. The normalized wet volume $V\phi/V_d$ is the inverse saturation. The times for TBP and TCP data have been rescaled in accordance to their viscosity $\mu$ and surface tension $\sigma$, as reflected in $D$. That is $t^{\prime}=t\frac{\mu^{\dag}}{\mu}\frac{\sigma}{\sigma^{\dag}}$ ($\mu^{\dag},\,\sigma^{\dag}$ are for TEHP liquid). This scaling is adequate for TBP, but TCP requires also recognition of its higher contact angle (as in Fig.4). The solid lines are power-law fits to the respective data; all, including the data for which the fit lines are not shown, conform to $V\sim t^{0.74\pm0.05}$.}\label{Fig3}
\end{figure}

The results of seven tests are summarized in Fig. \ref{Fig3}. All were taken at full symmetry, with various liquids and amounts as quoted in the figure. Inside the SFD regime (marked on the figure) all data fall neatly on a power law with exponent $0.74\pm0.05$. The small amount of scatter within each run suggests that the slight parallel shifts seen between the individual runs are due to small systematic errors such as deviations in the void fraction and liquid mass values used in the normalization. An estimated $\pm 15\%$ error in both these values would result in the error bar shown in Fig. \ref{Fig3}. As expected, the SFD regime is attained at $\bar{s}\sim10\%$, while deviations from the power law set in at a saturation of $\bar{s}\sim1\%$, and a limiting value of $\bar{s}\sim0.5\%$ is reached asymptotically over a period of many days. This may be due to the gradual decrease of active contacts between the grains (failure to form bridges due to insufficient liquid), which effectively diminishes the area ratio $S_e/S$ in (\ref{New1}). For an entire run signal intensity is rather uniform with individual grains appearing illuminated as hypothesized in the theoretical development. Additional tests carried out at half-symmetry have confirmed that in the SFD regime wetted volumes were hemispherical (Fig. \ref{Fig3}, inset) and followed the same time law.

\begin{figure}[htp]
\begin{center}
\includegraphics[trim=0cm 0cm 0cm 0cm,width=0.9\columnwidth]{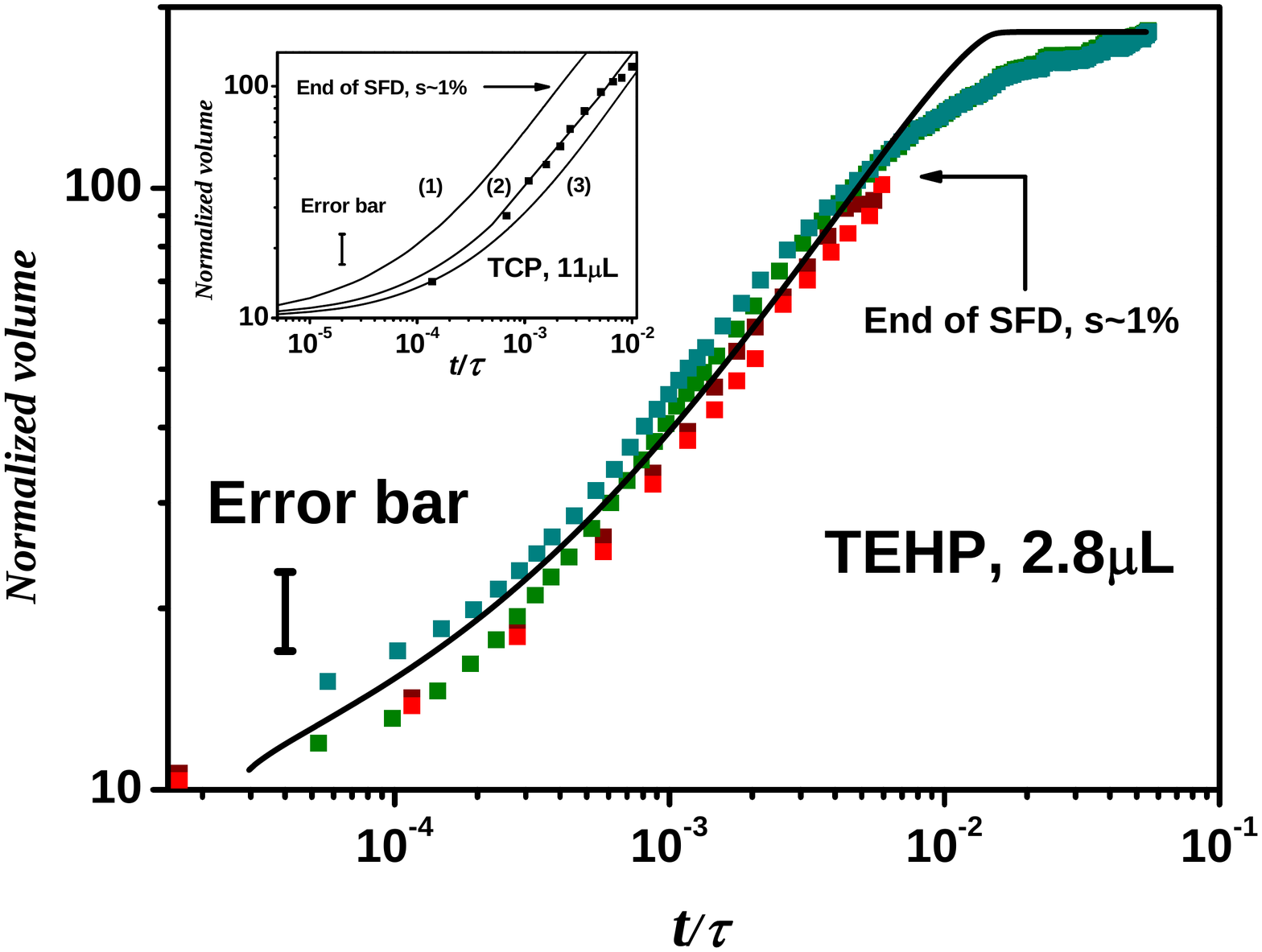}
\end{center}
\caption{A comparison of numerical results (solid lines) with the TEHP and TCP experimental data (symbols) and the structure of solutions to problem (\ref{NLDiff})-(\ref{frontV}) with varying $\tilde{s}_F$ (inset). The normalized wet volume $V\phi/V_d$ is the inverse saturation. The time is normalised by $\tau$ and is counted from the moment when $\bar{s}=10\%$. For TEHP/TCP, $D$ is $9.8\times10^{-15}/5.5\times10^{-15}\, \mbox{m}^2/\mbox{s}$. The numerical solutions were obtained at $s_F=0.55\%$ and $\tilde{s}_F=1.65\times10^{-2}$  (main plot) and (1) $\tilde{s}_F=5.5\times10^{-3}$, (2) $\tilde{s}_F=2.2\times10^{-2}$, (3) $\tilde{s}_F=3.85\times10^{-2}$ (inset).} 
\label{Fig4}
\end{figure}

To compare experimental results with what is predicted by our SFD model, problem (\ref{NLDiff})-(\ref{frontV}) has been solved numerically at 
$\gamma=-0.5$ assuming spherical symmetry by a semi-implicit moving-mesh method [9]. 
The basic character of the solutions is illustrated in Fig \ref{Fig4} (inset). 
We start simulations at the entrance to the SFD regime with a nearly flat initial saturation profile ($\bar{s}=10\%$, $\max(s)=15\%$ ); the initial wet area provides the length scale $L$ and $s_m=10\%$. 
At high/low capillary suction (low/high $\tilde{s}_F$) the spreading goes faster/slower at early times, but in all cases it develops into the $3/4$-power law seen in the experiments ($\pm0.02$). 
The effect of large/small value of $s_F/s_m$ is to move down/up the final wet volume. The experimental data for the $2.8\,\mu\mbox{L}$ TEHP drops were placed in Fig. \ref{Fig4} by measuring time from the moment when $\bar{s}=10\%$ and scaling it with $\tau$ calculated at $D=9.8\times 10^{-15}\,\mbox{m}^2/\mbox{s}$ and $L=3.6\,\mbox{mm}$. The numerical solution has been obtained by selecting $s_F=0.55\%$, according to the experimental data, and $\tilde{s}_F=1.65\times10^{-2}$. 
The unique combination of $D$ and $\tilde{s}_F$ has been found by variations of parameter $\delta$. 
This is possible since both $D$ and $\tilde{s}_F^{-1}$ are monotonic increasing functions of $\delta$ and, as one can see from Fig. \ref{Fig4} (inset), they both result in a parallel shift of the evolution curve in the same direction (the smaller/larger $\delta$, the smaller/larger both $D$ and $\tilde{s}_F^{-1}$, so the shift is to the right/left). 
The effective roughness amplitude ("film" thickness) for the TEHP experiments was found to be $\delta\approx140\,\mbox{nm}$ (with $N=8$). Another case chosen for comparison is the evolution of an $11\,\mu\mbox{L}$ TCP drop, which as may be noted stands alone in Fig. \ref{Fig3}. The data and numerical solution are shown in the inset to Fig. \ref{Fig4}. 
The TCP case requires $D=5.5\times 10^{-15}\,\mbox{m}^2/\mbox{s}$ and somewhat larger $\tilde{s}_F=2.2\times10^{-2}$ (smaller capillary pressure at the front) corresponding to $\delta\approx110\,\mbox{nm}$ (with $N=8$). This is consistent with the higher values of the contact angle found for this liquid. 
The obtained non-dimensional characteristic time scale $t/\tau\sim10^{-3}$, Fig. \ref{Fig4}, suggests that the time $t$ should be actually scaled by $s_F-s_r$ rather than by $s_m$. 
That is $\tau^{\prime}=\tau (\frac{s_F-s_r}{s_m})^{1-\gamma}\approx10^{-3}\,\tau$  substantiating the role of parameter $\tilde{s}_F$ and the capillary pressure in the process of spreading. 
The obtained values of $\delta$ are consistent with the effective "film" thickness estimate $\sim 100\,\mbox{nm}$ and suggest that the capillary action occurs in the finest range of the scales found in the roughness of the sand grains [11].

In conclusion, we have found that highly persistent liquids deposited on porous/granular media spread out to extremely low saturations in a transport regime not previously observable with normal liquids. The process is governed by an equation that belongs to the superfast diffusion class, and the result is a $3/4$ power law for the wetted volume as found in the experiments. The key physics is capillary-driven creeping flow within the roughness-defined channels, which are connected at the grain contacts by liquid bridges. The spreading stops at saturation $\sim0.5\%$, which is close to the level at which wet sand loses structural integrity. This suggests the crucial role of the bridges. The key parameter of the problem, $\tilde{s}_F$, expresses capillary suction at the wetting front and it is found to be related to the contact angle (surface energies). 

{\bf Acknowledgement.} This work is supported by the Joint Science and Technology Office,
Defense Threat Reduction Agency (JSTO/DTRA), Threat Agent Science (TAS) of the US Department of Defense. Special thanks to Dr. C.-H. Chang and Dr. S. Cai for the processing of MicroXCT data.

\vskip0.1truecm
\centerline{\rule{5truecm}{0.5pt}}
\vskip0.1truecm

\RefAIP{1}  M Scheel et al.   \yr{2008} \paper{Morphological clues to wet granular pile stability} \jour{Nature Materials} \vol{7} \pages{189--193}

\RefAIP{2}  M Scheel et al.   \yr{2008} \paper{Liquid distribution and cohesion in wet granular assemblies beyond the capillary bridge regime} \jour{J. Phys. Condens. Matter} \vol{20} \pages{494236}

\RefAIP{3}  S. Herminghaus \yr{2005} \paper{Dynamics of wet granular matter}  \jour{Advances in Physics} \vol{54} \pages{221-261}

\RefAIP{4} \by{Rye, R.R., Yost, F.G.,  O$^{'}$Toole, E.J.}  \yr{1998} \paper{Capillary Flow in Irregular Surface Grooves} \jour{Langmuir} \vol{14} \pages{3937--3943} 

\RefAIP{5} \by{Whitaker, S.} \yr{1969} \paper{Advances in Theory of Fluid Motion in Porous Media} \jour{Ind. Eng. Chem.} \vol{61} \pages{14-28}

\RefAIP{6} \by{Bico, J., Thiele, U., Quere, D.} \yr{2002} \paper{Wetting of textured surfaces} \jour{Colloid Surface A} \vol{206} \pages{41-46}

\RefAIP{7} \by {Orr, F.M., Scriven, L.E., Rivas, A.P.} \yr{1975} \paper{Pendular rings betweem solids: meniscus properties and capillary forces} \jour{\JFM} \vol{67} \pages{723--742}

\RefAIP{8} \by{J.L. Vazquez} {\em Smoothing and Decay Estimates for Nonlinear Diffusion Equations - 
Equations of Porous Medium Type} Oxford University Press \yr{2006}

\RefAIP{9} \by{M.J.Baines, M.E.Hubbard, P.K.Jimack} \yr{2011} \paper{Velocity-based moving mesh methods for nonlinear partial
 differential equations} \jour{ Comm.\ Comput.\ Phys.} \vol{10}  \pages{509--576}.

\RefAIP{10} \by{The Center for Research Information, Inc., 9300 Brookville Rd, Silver Spring, MD 20910} {\em Health Effects of Trioctyl Phosphate} \jour{Contract No. IOM-2794-04-001 of the National Academies}  \yr{2004}. D. G. Tuck,  Trans. Faraday Soc. {\bf 57}, 1297, (1961). R. D. Deegan, R. L. Leheny, N. Menon, S. R. Nagel, and D. C. Venerus, J. Phys. Chem. B {\bf 103}, 4066, (1999).

\RefAIP{11} \by{Alshibli, K.A., Alsaleh, M.I.} \yr{2004} \paper{Characterizing Surface Roughness and Shape of Sands Using Digital Microscopy} \jour{J. Comput. Civil Eng.} \vol{18} \pages{36--45}

\end{document}